\documentclass[CRPHYS,Unicode,manuscript]{cedram}

\usepackage{amsmath}
\usepackage{wasysym}

\def\horparallel{ \lower.5ex\hbox{ \includegraphics[width=2ex]{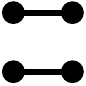}}\,\, }
\def\vertparallel{ \lower.5ex\hbox{ \includegraphics[width=2ex]{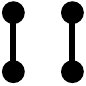}}\,\, }

\title{Classical and quantum spin liquids}

\alttitle{Liquides de spins classiques et quantiques}

\author{\firstname{Sylvain} \lastname{Capponi}\CDRorcid{0000-0001-9172-049X}}
\address{Laboratoire de Physique Th\'eorique, Universit\'e de Toulouse, CNRS, UPS, Toulouse, France}
\email[S. Capponi]{sylvain.capponi@univ-tlse3.fr}

\begin{abstract}
    When considering magnetic systems in the thermodynamic limit and at low enough temperature, one finds typically magnetically ordered phases. In contrast, in the high-temperature regime, the interactions between the spin degrees of freedom become less relevant and the system loses its order: this is a paramagnet. 
    This phenomenon of phase transition has been well understood using statistical mechanics and simple modelling. 

    In this short lecture notes, we will review the possibility that a many-body magnetic system may remain magnetically disordered down to zero-temperature, both for classical or quantum spins. These exotic phases of matter are known, respectively, as classical and quantum spin liquids. 

    We will address in particular the question of classification of these classical or quantum disordered phases. Indeed, while they have no local order parameter by definition, they can still possess different qualitative features related e.g. to the nature of their correlations or elementary excitations, which could be probed experimentally. 
    
\end{abstract}
\begin{altabstract}
   Lorsque l'on considère des systèmes magnétiques dans la limite thermodynamique et à   suffisamment basse température, on trouve des phases généralement ordonnées sur le plan magnétique. Au contraire, à haute température, les interactions entre les degrés de liberté de spin deviennent moins pertinentes et le système perd son ordre : c'est une phase paramagnétique. 
    Ce phénomène de transition de phase a été bien compris grâce à la mécanique statistique et à des modèles simples. 
    
    Dans ces brèves note de cours, nous examinerons la possibilité qu'un système magnétique  puisse rester désordonné magnétiquement jusqu'à la température nulle, à la fois pour des spins classiques ou quantiques. Ces phases exotiques de la matière sont connues respectivement sous le nom de liquides de spin classiques et quantiques. 

    Nous aborderons en particulier la question de la classification de ces phases désordonnées classiques ou quantiques. En effet, bien qu'elles n'aient pas de paramètre d'ordre local par définition, elles peuvent néanmoins posséder différentes caractéristiques qualitatives liées par exemple à la nature de leurs corrélations ou de leurs excitations élémentaires, qui peuvent être sondées expérimentalement. 
\end{altabstract}

\begin{document}
\maketitle

\section{Introduction}

Magnetism is a collective phenomenon that has been known and studied for extremely long time. Usual magnets are a direct evidence that a collection of microscopic elementary spin degrees of freedom tend to order along the same direction at low-enough temperature, while they are in random directions at higher temperature: this is the famous ferromagnetic transition. 
This phenomenon of phase transitions is well understood and can only occur in the thermodynamic limit, hence we will consider infinite systems. 
For simplicity, we will focus on localized spins and will not consider itinerant magnetism, where charge degrees of freedom of the carriers also play a role. 
Quite interestingly, simple models based on two-body interactions are enough to generate various properties, as is well-known in many-body physics. As was emphasized by Phil Anderson, "more is different"~\cite{Anderson1972}, which means that, within our simple framework, many-body physics can lead to various emergent properties: magnetic order, quasi-particles, gauge structure etc. which cannot be understood at the single-particle level. 

This review will focus on so-called "spin liquids", dubbed by analogy with the liquid phase of matter that does not have any magnetic order or does not break any symmetry. Still, we will see that this is a much richer concept since some spin liquids can nevertheless possess some structure giving rise e.g. to algebraic spin correlations or fractionalization of elementary excitations. In a modern formulation that we will explain, spin liquids can be viewed as fractionalized phases described by matter (\emph{spinons}) coupled to emergent gauge fields.

In a broader perspective, some features are linked to field theory, quantum many-body physics, quantum information, as well as experimental studies on several materials. We will point out some of these connections and refer to complementary reviews on these topics. For instance, an experimental definition of spin liquid is usually made by measuring the ordering temperature $T_c$  and computing the frustration parameter~\cite{Ramirez1994} $f=|\theta_{CW}|/T_c$ where $\theta_{CW}$ is the Curie-Weiss temperature (proportional to the spin exchange energy scale). Clearly a large ratio $f$ (typically $>10$) indicates that the ordering occurs at a much lower temperature than expected, pointing to a physical mechanism that prevents magnetic ordering. This usually occurs in frustrated materials with competing interactions, which is a very active area of research~\cite{LacroixMendelsMila_book,review_QSL_2017}. 

\begin{figure}
    \centering
    \includegraphics[width=0.35\linewidth]{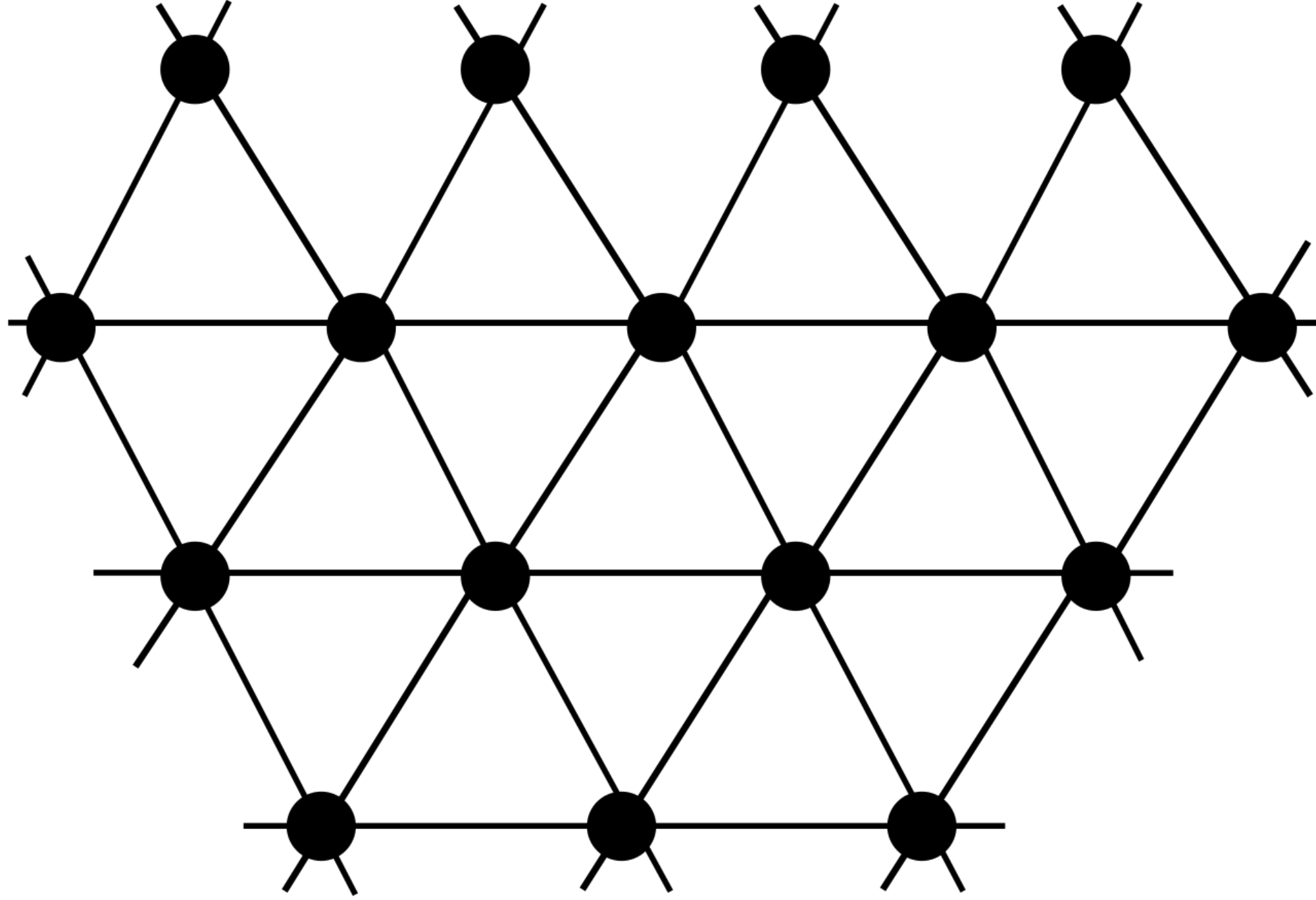}
        \includegraphics[width=0.35\linewidth]{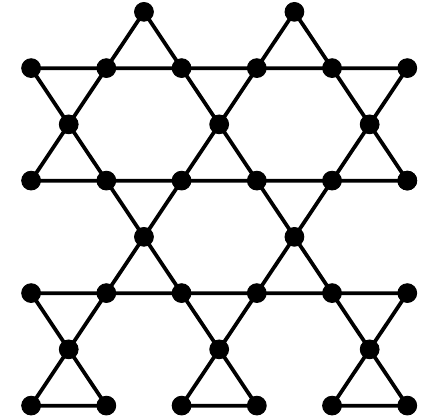}
         \includegraphics[width=0.35\linewidth]{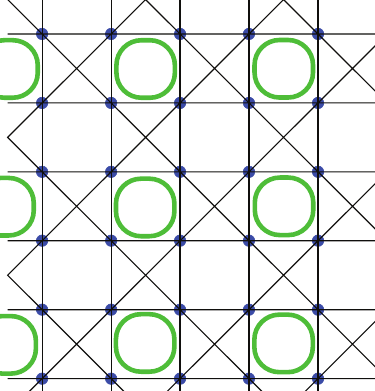}
        \includegraphics[width=0.35\linewidth]{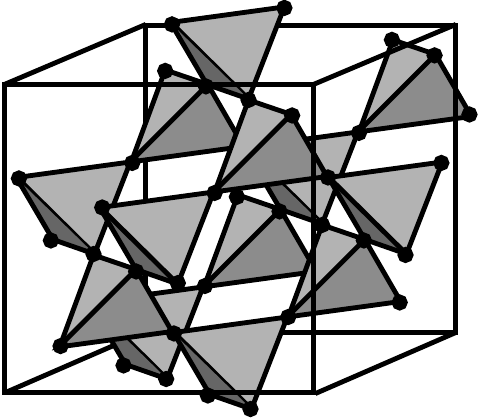}
    \caption{Structures of some lattices. From top left to bottom right: two-dimensional (2d) triangular, kagome, checkerboard and three-dimensional (3d) pyrochlore. For the checkerboard lattice, we have highlighted in green one plaquette phase groundstate in the quantum $S=1/2$ case.}
    \label{fig:kagome_pyro}
\end{figure}

In order to set up the stage, we will consider spin degrees of freedom, that can be discrete (Ising) or continuous (Heisenberg), localized on various regular lattices (square, honeycomb, triangular, kagome, pyrochlore etc.) in dimensions $d=1$, $2$ or $3$, see e.g. Fig.~\ref{fig:kagome_pyro}. We will also tackle constrained dimer models which are useful effective descriptions of some frustrated models at low-energy. 

As a disclaimer, we will not consider disorder, although it plays a crucial role in many situations, and will not discuss localization or spin glass behaviors.

In Sec.~\ref{sec:classical} and Sec.~\ref{sec:quantum}, we will treat respectively classical and quantum spin liquids. In both cases, we will try to define spin liquids, discuss possible classifications and argue that they do exist in some simple models. Although exact solutions are rare, they pave the way for investigating exotic phases of matter in a more realistic microscopic model as well as in real materials, which gives a support for the realization of classical and quantum spin liquids in nature.

\section{Classical spin liquids}\label{sec:classical}

A major success of statistical physics in the last century was to characterize the existence of a phase transition in the two-dimensional (2d) Ising model and related ones, showing that spins can order at low-temperature into a ferromagnet or antiferromagnet. 

Then in 1950, Wannier considered classical Ising spins on the frustrated triangular lattice~\cite{Wannier1950} and showed that there are an extensive number of groundstate configurations, contradicting the Nernst's principle of thermodynamics: this was the birth of classical spin liquids (CSL).

By definition, a CSL has an extensive degeneracy for its groundstates, which allows cooperative fluctuations and the absence of order. Of course such a situation is fragile and often unstable to fluctuations: this is the famous order-by-disorder mechanism~\cite{Villain1980}. Both thermal or quantum fluctuations will generally select an ordered state. Nevertheless, we do consider CSL since there can exist a large parameter regime where a cooperative paramagnet is the correct picture and the system remains magnetically disordered.~\cite{Villain1979}

We will now review some simple classical models where CSL can be stabilized. The classical variables can be either discrete (Ising spins) or continuous O(3) variables (Heisenberg spins) or even dimer configurations. For a more extensive review on this topic, we refer for instance to~\cite{Chalker2017}.

\subsection{Ising model}\label{sec:Ising}
The Ising model is one of the most famous statistical mechanics problem~\cite{review_Ising}. The energy of a configuration is simply given by 
\begin{equation}
    E(\{\sigma\}) =  J \sum_{\langle ij\rangle} \sigma_i \sigma_j
\end{equation}
where the sum runs over all nearest neighbor bonds of a lattice and $\sigma_i = \pm 1$ is the Ising spin variable at each site. 

This model has been very fruitful in the modern understanding of phase transitions: in one-dimension (1d), there is no transition; in 2d, there is the famous exact Onsager's solution; in 3d, the transition has been shown to be continuous but it is not yet proven whether it is conformally invariant. 

\begin{figure}
    \centering
    \includegraphics[width=0.5\linewidth]{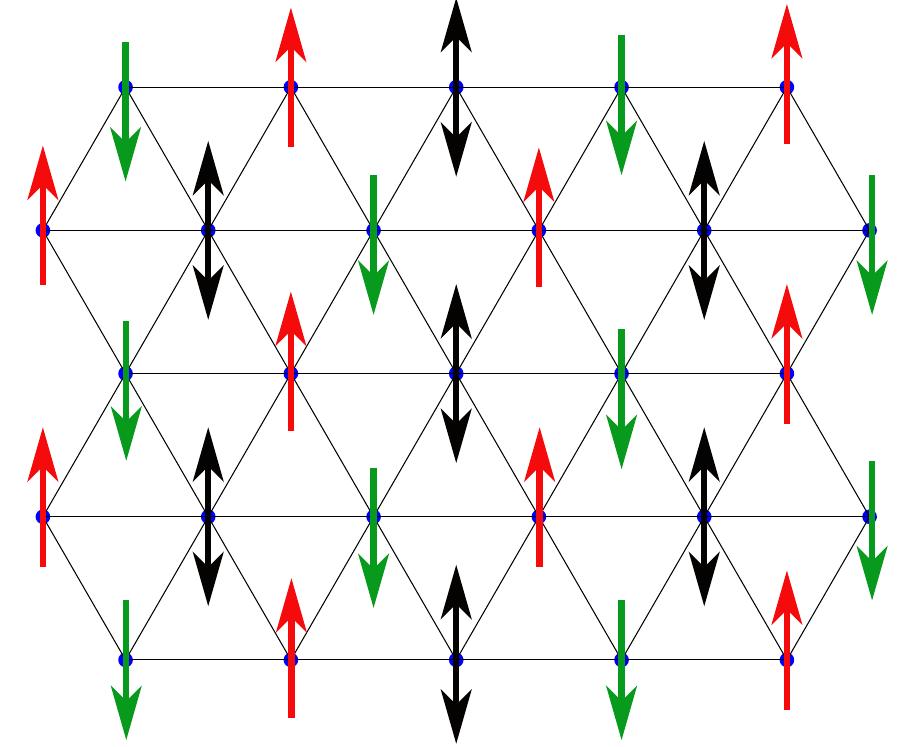}
    \caption{Triangular Ising antiferromagnet. Starting from a perfect N\'eel order on some honeycomb lattice (spins in red and green), the additional spins can be in any $\sigma=\pm 1$ state, leading to an extensive degeneracy.}
    \label{fig:triangular_isingAF}
\end{figure}

We have chosen the antiferromagnetic (AF) case ($J>0$) where frustration can lead to a large number of groundstates. Frustration means that no configuration can satisfy all constraints of having antiparallel spins on every bond. This has been analyzed by Wannier in 1950 on the triangular lattice~\cite{Wannier1950}, see Fig.~\ref{fig:triangular_isingAF}. Using a mapping to a dimer counting problem on the honeycomb lattice, it is possible to show that the residual entropy at zero temperature is $S_0/N \simeq 0.323 k_B$, which implies that that the number of groundstate configurations is extensive.

It is quite easy to get a rigorous bound by considering a perfect N\'eel AF on the non-frustrated bipartite honeycomb lattice with $N$ sites, see Fig.~\ref{fig:triangular_isingAF}. Then the additional sites of the triangular lattice can be in any $\sigma=\pm 1$ state, so that there are at least $2^{N/3}$ degenerate groundstates, i.e. the residual entropy per spin is larger than $k_B \log(2)/3 \simeq 0.231 k_B$.

Such a $T=0$ residual entropy violates the Nernst's third principle of thermodynamics and is common to many classical models. This Nernst's principle is more robust and obeyed in quantum mechanics although there are subextensive cases with fractons~\cite{Shirley2018} and even extensive degeneracy e.g. in the SYK model~\cite{SYK}.

Coming back to the triangular lattice case, there is no order at any finite temperature and correlations are algebraic~\cite{Chalker2017}, which appears quite peculiar a priori. This can be understood using e.g. mapping to constrained model as explained in the next section.

Quite interestingly, there are still several open questions when  considering additional further neighbor interactions, e.g. on the kagome lattice. There is no more any exact solution and the standard Monte-Carlo numerical simulations suffer from slowing down due to the large number of low-energy states. However, since the partition function can be written as a tensor network contraction, recent tensor algorithms have allowed to get a numerical solution.~\cite{Colbois2022}

\subsection{Constrained models}\label{sec:constrained}
\subsubsection{Vertex models}
\begin{figure}
    \centering
    \includegraphics[width=0.5\linewidth]{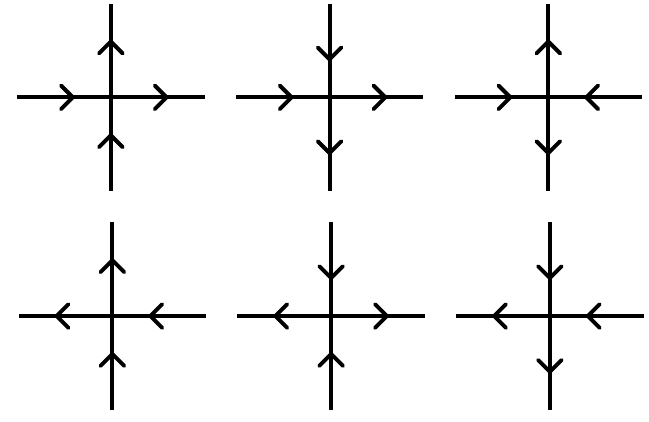}
    \caption{The six allowed configurations on the square lattice such that at each vertex, there are exactly 2 incoming and 2 outgoing arrows (ice rule).}
    \label{fig:sixvertex}
\end{figure}

One can also consider situations where the "spin" variables live on edges of the lattice, as done in lattice gauge theory, while interactions are on the vertices. In this framework, constraints can be introduced. For instance, on the square lattice, one can impose the so-called ice rule with a constraint of having  2-in/2-out arrows at each vertex, see Fig.~\ref{fig:sixvertex}. This is the famous six-vertex model solved by Baxter~\cite{Baxter_book}.

A simple estimate of the number of configurations for $N$ sites can be found using Pauling's estimate~\cite{Pauling1935}: since there are only 6 valid configurations around each vertex, instead of $2^4=16$, if one neglects correlations, there are approximately $2^{2N} (6/16)^N$ states, which leads to a residual entropy per site
\begin{equation}
    S_0/N = k_B \log(3/2) \simeq 0.405 k_B
\end{equation}
in good agreement with the exact solution by Lieb~\cite{Lieb1967}: $S_0/N = 3/2\log(4/3)k_B \simeq 0.431 k_B$. 
Note that a similar counting can be done on a three-dimensional (3d) pyrochlore lattice which has the same coordination number, see Fig.~\ref{fig:kagome_pyro}.

Such systems are called "ice" because there is a similar constraint for H atoms in water-ice and this residual entropy has indeed been measured experimentally long time ago~\cite{Giauque1936}. 

Quite interestingly, this local constraint leads to an effective gauge theory, similarly to Gauss' law in electromagnetism: charge conservation is expressed with $\mathrm{div} {\bf E} = 0$. As a result, using 
this analogy with electromagnetism, it can be shown~\cite{Chalker2017} that the system has critical (power-law) dipolar
correlations, leading to specific signatures, known as pinch points, in the structure factors: this is called a \emph{Coulomb phase}. 

In magnetism, these properties can be found in so-called spin ice materials, which have been quite popular recently as examples of  cooperative paramagnets having a Coulomb phase~\cite{Bramwell2020} and in which defects can be viewed as magnetic monopoles~\cite{Castelnovo2008}.

\subsubsection{Dimer models}
It is also possible to consider constrained models by putting hardcore dimers on all bonds of a lattice so that each site belongs to one and only one dimer (fully packed dimer configurations). This situation is very analogous to the previous vertex models, which can be seen as a constrained model with exactly two dimers per site. 
The number of configurations was solved in general on planar graphs~\cite{Kasteleyn1961,TemperleyFisher1961}. On bipartite lattices, it is generally possible to write down an effective field theory in terms of an height field, from which one can deduce that 
dimer correlations are dipolar. Such a phase is again a \emph{Coulomb phase}~\cite{Chalker2017}.

\subsection{Continuous spins: Heisenberg}

In some cases, a more appropriate description is provided by considering spins as classical $n$-component vectors of fixed length $S$. The simplest antiferromagnetic Heisenberg model is given by:
\begin{equation}
    {\mathcal H} = J \sum_{\langle ij\rangle} {\bf S}_i \cdot {\bf S}_j
\end{equation}
where $J>0$ is the AF exchange energy and the sum runs over nearest neighbors of the lattice.

On a Bravais lattice, by going to Fourier space, one finds that groundstate configurations correspond to the minimum of the Fourier transform $J({\bf q})$. For instance, on a triangular lattice, one finds a unique groundstate (up to symmetries) with a 120-degree spiral order.

When considering other non-Bravais frustrated lattices, the situation becomes more involved since in Fourier space $J({\bf q})$ is an $m\times m$ matrix, where $m$ is the number of spins per unit cell. In principle, one should find the smallest eigenvalue, keeping in mind that each spin has a fixed length: $\forall i\,\, |{\bf S}_i|^2=S^2$ (strong constraint). Since this cannot be done easily, the Luttinger-Tisza method~\cite{Luttinger1946} consists in assuming a \emph{weak constraint}:  $\sum_i |{\bf S}_i|^2 = NS$ which can provide a solution but not always.

On some lattices such as 2d kagome or checkerboard, or 3d pyrochlore, see Fig.~\ref{fig:kagome_pyro}, made of corner-sharing triangles or tetrahedra respectively, the classical Heisenberg model takes a simpler form:
\begin{equation}\label{eq:H_classical2}
 {\mathcal H} = \frac{J}{2} \sum_p {\bf \mathcal S}_p^2,
 \end{equation}
up to a constant, where ${\bf \mathcal S}_p$ is the total spin on a plaquette $p$ and the sum runs over all plaquettes.
From this expression, it is clear that if a configuration can satisfy ${\bf \mathcal S}_p={\bf 0}$ for all plaquettes, then it is a groundstate. Hence, one can look for solutions using so-called Maxwellian counting originally applied in mechanics of rigid bodies~\cite{Maxwell1864}. Following Chalker's presentation~\cite{Chalker2017}, we have $F=N(n-1)$ degrees of freedom for $n$-component spins. Let $N_p$ be the number of corner sharing units (plaquettes), each made of $q$ spins, hence  $N=qN_p/2$. In order to have a zero-energy configuration, one has to satisfy one constraint per cluster (total spin ${\bf \mathcal S}_p={\bf 0}$), i.e. $K=n N_p$ scalar constraints. Assuming that they can be satisfied and are linearly independent (which is obviously not true in general), one ends up with an effective number of degrees of freedom: 
$$\boxed{D=F-K = (\frac{q}{2}(n-1)-n)N_p}$$
For instance, when considering Heisenberg spins ($n=3$) on 3d pyrochlore or 2d checkerboard lattices (both having $q=4$), one finds that $D=N_p$ is extensive, leading to a macroscopic degeneracy and CSL behavior~\cite{MoessnerChalker1998a}. Quite remarkably for instance, on 3d pyrochlore, the system remains disordered at all temperature~\cite{MoessnerChalker1998b}.

Note that this argument does not provide any information for the kagome lattice, for which the constraints are indeed not independent. A proper treatment leads to $D=N/9$ zero modes in this case and the order-by-disorder mechanism selects coplanar spin configurations~\cite{Chalker1992}. 
Finally, let us mention that frustration can also occur from competing interaction, e.g. next-nearest-neighbor coupling $J_2$ on the square lattice: this is the famous $J_1$-$J_2$ Heisenberg model on the square lattice, which also has a very large degeneracy when $J_2=J_1/2$.

\subsection{Classification}
Inspired by the Luttinger-Tisza model, a classification can be made depending on the structure of the flat bands in the spectrum. Indeed, for the particular case (\ref{eq:H_classical2}), the hamiltonian can be brought in this form:
\begin{equation}
 {\mathcal H} = \frac{J}{2} \sum_{\ell,m} \sum_q (L^\ell_q L^m_{-q}) S^\ell_q \cdot S^m_{-q}
\end{equation}
defining an $n$-component vector ${\bf L}(q)$, from which the number of dispersive modes can be obtained.~\cite{Davier2023,Yan2023}

Quite interestingly, the analysis of ${\bf L}(q)$ allows to get higher-rank tensor, e.g. Gauss' law with tensors with fracton excitations, with quadratic or quartic dispersion relations etc.
Analysing the band structure, the gap closing points, the number of pinch points and their nature allow to classify different CSL~\cite{Davier2023,Yan2023}.

Closing this section on CSL, it is very exciting to see the ongoing classification as well as novel exotic CSL (algebraic, fracton etc.) which could potentially be relevant in real materials. It would be interesting to go beyond the specific case (\ref{eq:H_classical2}) for a complete classification of CSL.

Last but not least, CSL are parent states when quantum fluctuations start playing a role and natural starting point to realize exotic quantum states.

\section{Quantum spin liquids}\label{sec:quantum}
\subsection{General features}
There are different ways to add quantum fluctuations in a classical model, for instance adding a transverse magnetic field or XY exchange to an Ising model, or simply promoting the spins into quantum operators in the Heisenberg model. Based on the order-by-disorder mechanism, one expects on general grounds that quantum fluctuations will select some ordered phase. In this part, we will focus on groundstate properties at zero temperature in quantum models and review the possibility to get unconventional phases of matter, in particular quantum spin liquids (QSL) that we will define. 

Note that the definition is often negative, in the sense that such a QSL phase does not break any symmetry, but recent advances have shown the role of several observables: quantum entanglement, topology as well as spectroscopic features~\cite{Savary2017}, which could help to characterize and distinguish various QSL. 

\begin{figure}
\begin{center}
\includegraphics*[width=0.4\linewidth]{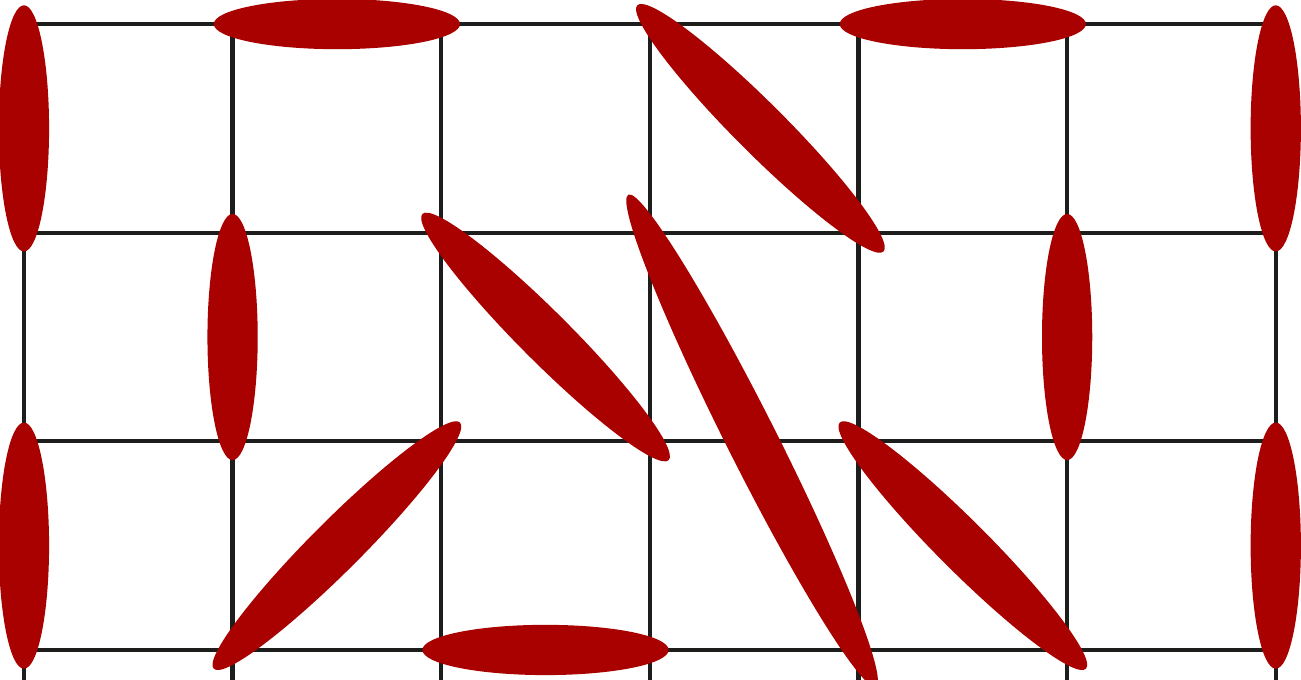}
\end{center}
\caption{Example of a valence-bond configuration on the square lattice. Each dimer corresponds to a singlet state made of two spins-1/2. An equal weight superposition of all coverings is known as the RVB wavefunction.}
\label{fig:rvb}
\end{figure}

The story of QSL goes back to Anderson's idea of a resonating valence-bond (RVB) groundstate, that is a coherent superposition of valence-bond configurations (made of 2-site singlets of spin 1/2)~\cite{Anderson1973}, see Fig.~\ref{fig:rvb}. By definition, such a state does not break any spin or lattice symmetries, as expected for a spin liquid. 
From a variational point of view, a nearest-neighbor RVB state (made only of nearest-neighbor singlet) has a variational energy per site of $(-3/8)J$ which is already lower than the one of a classical N\'eel state ($-(1/4)J$) on a spin-1/2 Heisenberg chain. Hence it was proposed to be a good candidate for $S=1/2$ Heisenberg model on the triangular lattice, which turned out to be wrong since this model is now believed to be ordered magnetically. But it has been a fruitful idea ever since and we will discuss some exotic properties of this wavefunction as well as its relevance for some simple microscopic models.

\begin{figure}
\begin{center}
\includegraphics*[width=0.4\linewidth]{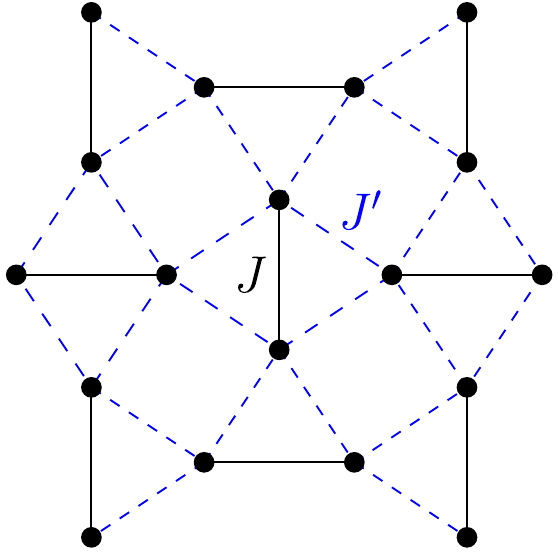}
\end{center}
\caption{Shastry-Sutherland lattice for which the product of singlets on the $J$ bonds is an exact eigenstate and the unique groundstate for small enough $J'/J$.}
\label{fig:shsl}
\end{figure}

On the one hand, there exist some simple groundstates that obviously do not break spin or lattice symmetries, such as a trivial singlet product state on the Shastry-Sutherland lattice, see Fig.~\ref{fig:shsl}. Such a product-state is featureless and does not possess any exotic property.~\cite{Shastry-S-1981} 

On the other hand, we know that in condensed matter systems, some gapped phases can be nontrivial in the sense of having topological order, e.g. the famous fractional quantum Hall effect (FQHE). Topological phases of matter are gapped quantum phases 
containing nontrivial features which are not due to spontaneous
symmetry breaking. While such 
 systems have exponentially decaying correlation
and look quite simple from a classical point of view,
they do exhibit nontrivial properties. For example, the groundstate
degeneracy can depend on the topology of the closed
manifold (torus vs sphere), or there can exist protected gapless
edge excitations if the system has a boundary, or bulk excitations can possess nontrivial statistics etc. Generally also, there are  nontrivial features in the bipartite entanglement spectrum~\cite{LiHaldane2008}. Clearly, QSL are much richer than their classical analogue, and the role of quantum order was pointed out by Wen~\cite{Wen2002}.

Since QSL cannot be classified according to symmetry breaking, one possible classification was proposed based on the nature of the elementary excitations (gapped or gapless spinons) as well as the emergent gauge field (U(1), SU(2), Z$_2$\ldots) which mediate their interactions~\cite{WenBook}.
More recently, entanglement properties were also shown to be relevant for this classification: quantum spin liquids are by definition quantum states that are not connected adiabatically to trivial product states~\cite{Savary2017}. There are two classes: (i) 
long-range entangled (LRE) if it cannot be adiabatically connected to a product state under any local unitary transformation (e.g. FQHE, Z2 QSL, chiral spin liquid etc.); (ii) 
short-range entangled (SRE) if it cannot be adiabatically connected to a product state while respecting some symmetries: symmetry protected topological (SPT) state, e.g. Haldane $S=1$ chain, topological insulator etc.

There are intimate relations between these definitions since topological order is probably needed to realize fractionalization in dimension $d\geq 2$~\cite{OshikawaSenthil2006}.

Building on RVB idea, Kalmeyer and Laughlin have proposed a chiral spin liquid phase to be realized on the triangular lattice~\cite{Kalmeyer1987}. This phase is analogous to the FQHE on a lattice: existence of chiral edge modes, quantized thermal transport, bulk excitations are anyons etc. We will see later microscopic models where it is possibly realized.

\subsection{Classification based on many-body spectrum}
We will now consider the low-energy features of the many-body spectrum of a generic model hamiltonian ${\mathcal H}$. Note that sometimes we only have a wavefunction (e.g. RVB), but it is possible to consider a parent Hamiltonian for which this is the groundstate, see later. For the sake of simplicity, we will assume ${\mathcal H}$ to be short-ranged and local, most of the time with some U(1) or SU(2) spin symmetry. A typical example is the Heisenberg model
\begin{equation}
    {\mathcal H} = J \sum_{\langle ij\rangle} \hat{S}_i \cdot \hat{S}_j + \ldots
\end{equation}
where $J>0$ is the AF coupling constant, $\hat{S}$ are quantum spin operators ($S=1/2$, $1$, \ldots) and additional interactions could be needed to stabilize a QSL groundstate.

\begin{figure}
\begin{center}
\includegraphics*[width=0.8\linewidth]{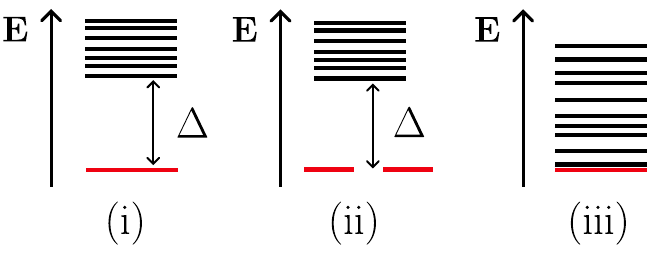}
\end{center}
\caption{Three different possibilities for the many-body spectrum: (i) a unique groundstate and a finite gap $\Delta$; (ii) degenerate groundstates and a finite gap $\Delta$; (iii) gapless. }
\label{fig:spectrum}
\end{figure}

Focusing on low-energy, we can distinguish three qualitative different spectra, see Fig.~\ref{fig:spectrum}: (i) a unique groundstate and a finite gap $\Delta$; (ii) degenerate groundstates and a finite gap; (iii) gapless.
We will see in the following some examples, as well as some refinements regarding case (ii) since the degeneracy can result either from symmetry breaking or from the topological nature.

We will make use of exact and rigorous results, when available. Indeed, even if some models are fine-tuned or artificial, they have the merit of showing the existence of exotic quantum phases. For realistic and generic models, I will refer to several numerical studies, with the caveat that finite-size effects cannot be fully controlled. Indeed, it has been shown that deciding whether a 2d translation-invariant hamiltonian has a gap or not in the thermodynamic limit is undecidable~\cite{Cubitt2015}\ldots

When discussing the low-energy spectrum, a seminal result was provided by Lieb, Schultz and Mattis (LSM)~\cite{LSM1961} in 1d, which was later extended to arbitrary dimension $d$~\cite{Oshikawa2000,Hastings2004}. Basically,  
LSM forbids having case (i) spectrum for an odd number of half-integer spins per unit cell.

Note that a similar result can also be obtained in finite magnetic field~\cite{Oshikawa2000}, since LSM relies on interplay of U(1) symmetry and translations, constraining whether some magnetization plateaux phases can be featureless or not.~\footnote{A quantum groundstate is said to be featureless when it can be adiabatically connected to a trivial product state.} In any dimensions, a featureless magnetization plateau is possible iff $nS(1-m) =\mathrm{integer}$, where $n$ is the number of spins per unit cell, $S$ the spin value and $m$ the total magnetization normalized by its maximal value.

Let us mention some recent extensions that have been obtained when considering symmetries beyond translations (nonsymmorphic or point-group) or by matching UV/IR anomalies in field theory. Namely, the groundstate of a $S=1/2$ hamiltonian \emph{cannot} be featureless (case (i))~\cite{Liu2024}:
\begin{itemize}
\item on the diamond lattice\cite{Parameswaran2013a};
\item if there is an even-order rotation symmetry~\cite{Po2017};
\item on the pyrochlore lattice (4 spins per unit cell)~\cite{Ye2022}.
\end{itemize}

Note that when LSM-like argument does not apply, a featureless groundstate, depicted as case (i), is possible but it is not always straightforward to construct one. There are explicit constructions e.g. for $1/3$-magnetization plateau on the kagome lattice~\cite{Parameswaran2013b} or a featureless spin-1/2 wavefunction on the honeycomb lattice~\cite{Kimchi2013,Kim2016}.

\subsection{Classification based on slave particle representations}
In order to make some analytic progress, it can be convenient to use exact fermionic or bosonic representations of spin operators, known as slave particles due to some constraints.~\cite{Lee2006}

For instance, one can introduce Abrikosov fermions
\begin{equation}
    \hat{S}^z = \frac{1}{2}(c^\dagger_\uparrow c_\uparrow - c^\dagger_\downarrow c_\downarrow)\qquad S^+ = c^\dagger_\uparrow c_\downarrow \qquad S^- = c^\dagger_\downarrow c_\uparrow
\end{equation}
which is a faithful representation of a spin-1/2 at each site provided a single occupancy constraint is added. In this language, the spin-spin interaction is quartic so that a natural mean-field approximation can be performed, leading to 
\begin{equation}
    {\mathcal H}_\mathrm{MF} = \sum_{\langle ij\rangle} \chi_{ij} (c^\dagger_{i\uparrow} c_{j\uparrow} + c^\dagger_{i\downarrow} c_{j\downarrow}) + \eta_{ij} (c_{i\uparrow} c_{j\downarrow}-c_{i\downarrow}c_{j\uparrow}) + h.c.
\end{equation}
which is quadratic and can be solved. Of course, this mean-field solution may or may not be stable when considering fluctuations, in particular the gauge field enforcing the one fermion per site constraint. This is thus amenable to analytic studies or numerical ones when performing an exact Gutzwiller projection~\cite{Gutzwiller1963} to enforce the constraint. 

Similarly, a slave-boson description can be obtained from Schwinger bosons:
\begin{equation}
    \hat{\bf S}  = \frac{1}{2} b^\dagger_\alpha {\mathbf \sigma}_{\alpha \beta} b_\beta
\end{equation}
imposing a constraint $\sum_\alpha b^\dagger_\alpha b_\alpha=2S$ at each site. This also allows to perform mean-field decoupling. Note that from a numerical point of view, one needs to work with permanent rather than determinant for the projection, which is very costly. 

In all theses approaches, known as parton constructions, there is some redundancy of the mean-field descriptions, revealing the gauge nature of these theories. This has been used to classify possible solutions (Ansatz) using a projective symmetry group (PSG) approach~\cite{Wen2002}. In particular, different spin liquids that do not break any symmetry can have different PSG. Even though this is obtained from a mean-field approach, it should be a property of the phase itself.

Such a PSG classification is still being worked on in the community, for bosonic or fermionic spinons on all lattices. For instance, when time-reversal symmetry is broken (as in chiral spin liquid), an additional classification is needed~\cite{Messio2013,Bieri2016}.

Depending on the nature of the gauge fields and the spinon spectrum, several QSL could be realized~\cite{Lee2006}, e.g.:
\begin{itemize}
    \item For a Z$_2$ gauge field and gapped spinons, a gapped Z$_2$ QSL phase emerges, which is stable in 2d and 3d;
    \item For a U(1) gauge field and gapped spinons, a genuine QSL is unstable in 2d~\cite{Polyakov1977} towards a valence-bond crystal (VBC) phase that break some lattice symmetries;
    \item For a U(1) gauge field and gapless Dirac spinons, the situation is not fully settled and an algebraic Dirac spin liquid could be stabilized in $(2+1)$d~\cite{Hermele2004a}.
\end{itemize}

What is even more dizzying is that there could exist several different QSL within the same gauge symmetry class, dubbed symmetry-enriched topological (SET) order, which can be analyzed using topological quantum field theory~\cite{Barkeshli2019}. As an example, there are about $2^{21}$ different gapped Z$_2$ QSL on the 2d square lattice~\cite{Essin2013} !

This parton construction approach has been very fruitful to provide various QSL Ansatz and is still ongoing effort to classify QSL in 3d on various lattices, see e.g.~\cite{Chauan2023,Liu2024}.

\subsection{Unique groundstate and finite gap}
\subsubsection{Trivial phase}
Of course, in some situations a featureless paramagnetic groundstate is possible, with a finite gap to all excitations. It can be adiabatically connected to a trivial product state. For instance, this is the case for Heisenberg spin ladder or bilayers, including some exact results for fine-tuned frustrated models~\cite{LacroixMendelsMila_book}. Quite interestingly, there is also the famous example of Shastry-Sutherland lattice~\cite{Shastry-S-1981}, see Fig.~\ref{fig:shsl}, on which the groundstate is a product state of singlets for small enough $J'/J$. 

In such trivial phase, correlations are short-range and
there is no topological entanglement entropy (a subdominant contribution to the area law).
Note that it can still lead to rich physics when adding e.g. a finite magnetic field with the appearance of nontrivial magnetization plateaux as well as superfluid or supersolid phases~\cite{LacroixMendelsMila_book}.

\subsubsection{One-dimensional SPT phase}
A famous example of 1d SPT phase is provided by the Haldane phase of a $S=1$ Heisenberg chain~\cite{Haldane1988}, which can be understood from the exact groundstate of a slightly deformed hamiltonian known as AKLT model~\cite{AKLT}:
\begin{equation}\label{eq:AKLT}
    {\mathcal H}_\mathrm{AKLT} = J\sum_i \left( \hat{S}_i\cdot \hat{S}_{i+1} + \frac{1}{3}(\hat{S}_i\cdot \hat{S}_{i+1})^2\right).
\end{equation}
While the groundstate is unique on a chain with periodic boundary conditions (PBC), it becomes 4-fold degenerate in the presence of open boundary conditions (OBC) due to the emergence of $S=1/2$ edge states. Thus, it has some topological properties, which are also revealed in the bipartite entanglement spectrum. It has been understood that these phases are not adiabatically connected to trivial ones, provided some symmetries are present~\cite{Pollmann2012}, hence their name: symmetry-protected topological (SPT) phases.

\subsubsection{Two-dimensional SPT phase}
It turns out that AKLT construction~\cite{AKLT} can also be performed in higher dimension, e.g. $S=3/2$ on the honeycomb lattice or $S=2$ on the square lattice. For these models, it can be shown that the groundstate is unique and all correlations decay exponentially. While there is no rigorous proof of a finite-spin gap, all numerical studies point to a finite value, see e.g.~\cite{Lemm2020}.

Such states are also known as valence-bond solids (VBS) since they can be understood from valence-bond configurations and they do not break any spin or lattice symmetries. 

Similarly to the 1d case, the properties depend on having periodic vs open boundary conditions, which signals some topological properties: these states are called 2d SPT phases.~\cite{Chen2011b}

\subsection{Gapped spectrum with groundstate degeneracy}
\subsubsection{Spontaneous symmetry breaking}
The most well-known example of a gapped spectrum with groundstate degeneracy occurs when there is a spontaneous symmetry breaking (SSB) of a \emph{discrete} lattice symmetry. This is known to occur e.g. in the 1d Majumdar-Ghosh $S=1/2$ model~\cite{MajumdarGhosh1969}:
\begin{equation}
    {\mathcal H}_\mathrm{MG} = J\sum_i \left( \hat{S}_i\cdot \hat{S}_{i+1} + \frac{1}{2} \hat{S}_i\cdot \hat{S}_{i+2} \right) ,
\end{equation} where a spontaneous dimerization occurs, or in various 2d spin models with columnar or plaquette orders, e.g. $J_1$-$J_2$-$J_3$ $S=1/2$ on the honeycomb lattice~\cite{Fouet2001} or $S=1/2$ on the checkerboard lattice~\cite{Fouet2003} (see Fig.~\ref{fig:kagome_pyro} where one plaquette phase is sketched with green symbols) and so on.
Quite interestingly, a similar mechanism can also occur in the presence of magnetic field giving rise to magnetization plateaux with SSB, e.g. exact magnon states close to saturation in frustrated lattices~\cite{Schulenburg2002} or similar mechanism at other magnetization values~\cite{Capponi2013}.

These situations are well understood and magnetic excitations are conventional, e.g. gapped $S=1$ magnons in 2d.

\subsubsection{Topological phase}
Quite generally, gapped topological phases can also be characterized by the fact that elementary excitations are gapped anyons, which statistics can be nontrivial, e.g. nonabelian,  which could be useful for topological quantum computation. The groundstate degeneracy depends on the genus of the manifold, which is quite different from the usual spontaneous symmetry breaking. Remarkably, the fractionalization of excitations is strongly tied to the topological properties~\cite{Wen1991,OshikawaSenthil2006} and could be detected in broad continuum-like excitations in the spin dynamical structure factors $S({\bf q},\omega)$.

At a more fundamental level, a gapped topological phase in $(2+1)$d can be probed using the topological entanglement entropy~\cite{LevinWen2006,KitaevPreskill2006}, which is a subleading constant term to the usual area law.

\subsubsection{Some specific models}
In this endeavour to discover QSL in realistic models or materials, the quest for exact solution or controlled approximation is crucial. Indeed, this allows to show that some of these phases are possible, sometimes stable, and hence relevant to more generic situations.

\begin{figure}
\begin{center}
\includegraphics*[width=0.3\linewidth]{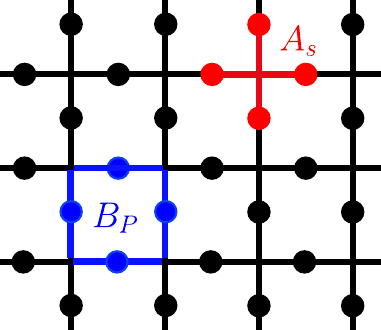}
\end{center}
\caption{Illustration of Kitaev's toric code model on the square lattice.}
\label{fig:kitaev}
\end{figure}

\paragraph{Kitaev's toric code}
In 2003, Kitaev has introduced the toric code as an exact model for which the groundstate is a gapped Z$_2$ topological phase~\cite{Kitaev2003} in $(2+1)$d. 
As in a gauge theory, the Ising spins live on the links of a square lattice and the Hamiltonian is given by 
\begin{equation}\label{eq:kitaev}
    {\mathcal H}_\mathrm{toric} = -\sum_s A_s - \sum_P B_P 
\end{equation}
where the first (respectively second) sum runs over all sites (respectively plaquettes) of a 2d square lattice, $A_s$ is the product of $\sigma_x$ on all links around a site $s$ while $B_P$ is the product of all $\sigma_z$ on all bonds of a plaquette $P$, see Fig.~\ref{fig:kitaev}.

Since all $A_s$ and $B_P$ commute together, an exact solution can be provided. The groundstate is featureless and there are four types of excitations (all with quantum dimensions $d_i=1$): trivial, $e$, $m$, $f=e-m$ pair. 
$e$ and $m$ have $\pi$-shift mutual statistics, hence they are anyons (abelian) and $f$-excitations behave as fermions. 

This solution was a breakthrough since it has shown the existence of a topological phase, analogous to the famous $\nu=1/3$ FQHE, in a simple spin model. However, it is very fine-tuned since the quasiparticles have no dispersion and the correlation length is zero.

\begin{figure}
\begin{center}
\includegraphics*[width=0.3\linewidth]{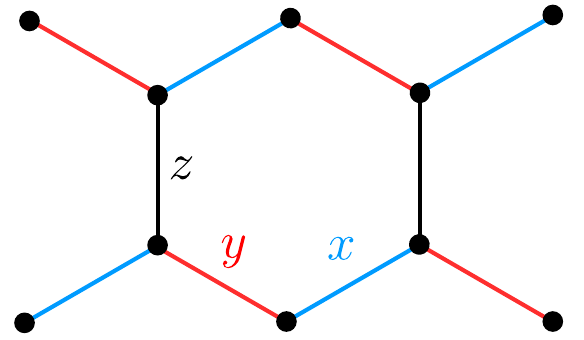}
\end{center}
\caption{Illustration of Kitaev's honeycomb model.}
\label{fig:kitaev2}
\end{figure}

\paragraph{Kitaev's honeycomb model}\label{sec:kitaev2}
Later, Kitaev has found an even more interesting exact solution on the honeycomb lattice~\cite{Kitaev2006}. The model is defined in terms of spin-1/2 variables on the sites of a 2d honeycomb lattice:
\begin{equation}\label{eq:kitaev2}
    {\mathcal H}_\mathrm{honeycomb} = -J_x \sum_x \sigma_j^x \sigma_k^x -J_y \sum_y \sigma_j^y \sigma_k^y -J_z \sum_z \sigma_j^z \sigma_k^z 
\end{equation}
where each term acts on the three different kind of bonds, see Fig.~\ref{fig:kitaev2}.

Using a faithful Majorana representation of the spin operators, the exact solution provides a gapped or gapless phase depending on the parameters. The gapped phase is similar to the previous toric code case, while the gapless phase is more intriguing: in the presence of a magnetic field, a gapped phase with nonabelian anyons can be stabilized, similar to a topological superconductor.

Following this seminal work, a generalization by Levin and Wen has given access to a huge zoo of topological phases in so-called string-net models~\cite{LevinWen2005}.

\begin{figure}
\begin{center}
\includegraphics*[width=0.4\linewidth]{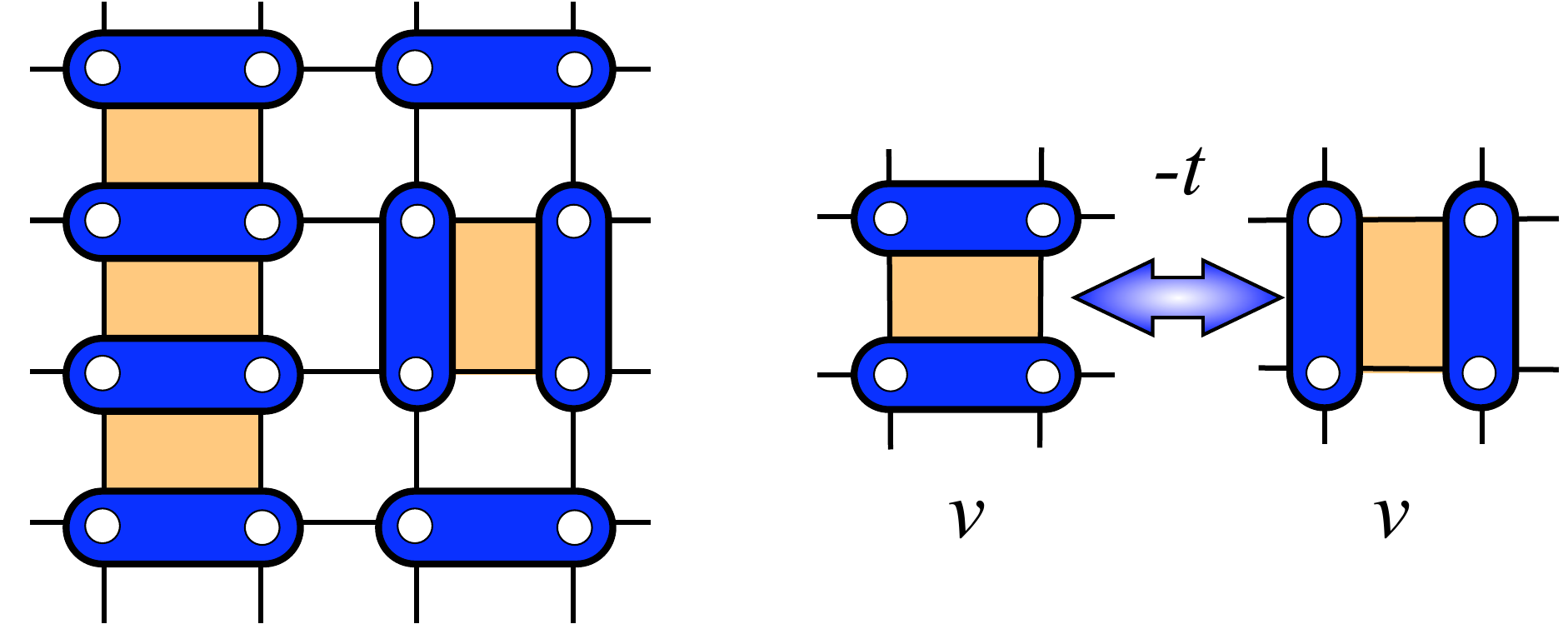}
\end{center}
\caption{Left: A dimer configuration on the square lattice. Flippable plaquettes are shaded.
Right: Flippable plaquettes contribute a diagonal term $v$ to the Hamiltonian and can be flipped with amplitude $-t$ $(t>0)$.}
\label{fig:QDM}
\end{figure}

\paragraph{Quantum dimer models}
In the context of QSL, quantum dimer models (QDM) have been very inspiring examples and we refer to the review by Moessner and Raman in Ref.~\cite{LacroixMendelsMila_book} for an extensive discussion. The Hilbert space consists in fully-packed dimer configurations (i.e. as in classical dimer models) and quantum fluctuations arise from a simple model introduced by Rokhsar and Kivelson (RK)~\cite{RokhsarKivelson1988}, e.g. on the square lattice:
\begin{align}
  \nonumber H_{\rm QDM}= \sum \Big[ & -t \left (\left| \horparallel
  \right\rangle\left\langle \vertparallel \right| + \rm{h.c.}\right) + \\ &
  \quad\quad v \left( \left| \horparallel \right\rangle\left\langle
  \horparallel\right| + \left| \vertparallel \right\rangle\left\langle
  \vertparallel \right|\,\right) \Big]\,,
  \label{rk-hamiltonian}
\end{align}
(in terms of hard core dimer objects $\vertparallel$) where $t$ and $v$ are the 
amplitudes of kinetic and potential terms, and
the sum runs over all elementary square plaquettes. Graphically, these processes are represented in Fig.~\ref{fig:QDM}.
As in the classical case, the local constraint provides a natural gauge structure and conserved quantities: winding number and U(1) symmetry on bipartite lattices (e.g. square), or topological sectors and Z$_2$ symmetry on non-bipartite ones (e.g. triangular).

\begin{figure}
\begin{center}
\includegraphics*[width=\linewidth]{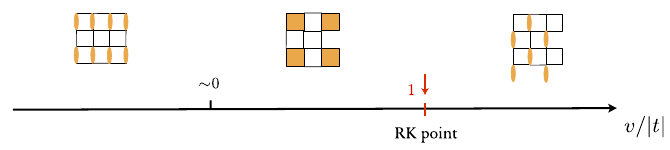}
\end{center}
\caption{Schematic phase diagram of the QDM on the square lattice as a function of $v/t$: the critical groundstate at the RK point is unstable to crystalline phases such as columnar, plaquette or staggered.}
\label{fig:qdm_square}
\end{figure}

Precisely when $v=t$, so-called RK point, the groundstate can be obtained as an equal weight superposition of all configurations. On most bipartite lattices, it is part of a Coulomb phase (with algebraic correlations) but unstable towards crystalline phases that break lattice symmetries, see Fig.~\ref{fig:qdm_square}. On nonbipartite lattices, the RK point belongs generically to an extended gapped Z$_2$ RVB topological phase~\cite{MoessnerSondhi2001}.

The instability of U(1) QDM is linked to a famous result by Polyakov about gauge theories~\cite{Polyakov1977}. However, Z$_2$ gauge theories can be deconfined in $(2+1)$d in $(3+1)$d as well as U(1) ones in $(3+1)$d.

As a conclusion, QDM models can exhibit various nonmagnetic phases including valence bond crystals (VBC), Coulomb phase or gapped Z$_2$ RVB topological phase.

\paragraph{Anisotropic spin models}
Following the success of QDM to describe unconventional phases, a natural roadmap was to engineer spin models with strong anisotropies (in spin space), such that QDM emerge as en effective description. 

For instance, let us mention a model introduced on the kagome lattice (see Fig.~\ref{fig:kagome_pyro}) by Balents, Fisher and Girvin~\cite{Balents2002}:
\begin{equation}\label{eq:BFG}
    {\mathcal H}_\mathrm{BFG} = J_z \sum_{\hexagon} (S^z_{\hexagon})^2 - J_\perp \sum_{\hexagon} \left\{(S^x_{\hexagon})^2 + (S^y_{\hexagon})^2\right\}
\end{equation}
where $J_z$ is the dominant energy scale that constrains the total $S_z$ to vanish on each hexagonal plaquette, i.e. there are exactly 3 up and 3 down spins per hexagon. As a result, the model maps at low-energy onto an effective QDM-like model on the dual triangular lattice, where there are exactly three dimers per site. Subsequent large-scale numerical studies have shown that this model does exhibit fractionalization and realizes a gapped Z$_2$ topological phase~\cite{ShengBalents2005}, which is adiabatically connected to the RK point of this QDM-like model.

Quite interestingly other spin models are amenable to large-scale unbiased quantum Monte-Carlo (QMC) simulations, e.g. an SO($N$)-symmetric spin model on the kagome lattice which groundstate is a gapped Z$_2$ QSL~\cite{Block2020}.

\paragraph{SU(2)-symmetric spin models}
In order to get closer to realistic spin models, similar physical properties were investigated in SU(2)-symmetric spin models, e.g. using decorated lattices to reproduce QDM physics~\cite{Raman2005}. As a result, it was possible to construct groundstate having gapped Z$_2$ or gapless U(1) properties in 2d and 3d spin models.

Later, Cano and Fendley have proposed SU(2) spin-1/2 models with local interactions that can stabilize RVB groundstates, similar to the RK points of the corresponding QDM model~\cite{Cano2010}. As discussed previously, depending on the lattice, it can be a quantum critical point or a genuine gapped spin liquid phase. Quite interestingly, by adapting this construction, it is also possible to engineer a spin model having similar properties as the QDM one for any interaction $v/t$ parameter~\cite{Mambrini2015}.

\paragraph{Chiral QSL}
We have already mentioned the original proposal by Kalmeyer and Laughlin~\cite{Kalmeyer1987} to realize in a spin system a quantum phase analogous to the $\nu=1/2$ bosonic FQHE phase. This is a gapped phase, with 2-fold degeneracy on a torus. With open boundaries, a rich edge physics emerges with gapless chiral modes described by an SU(2)$_1$ conformal field theory (CFT). 

Regarding microscopic models, several large-scale numerical studies have shown that a chiral QSL groundstate can be stabilized on various spin-1/2 models with local interactions on the kagome or triangular lattice with explicit time-reversal symmetry breaking~\cite{Bauer2014,Gong2014,He2014,Gong2015,Wietek2017}
or without~\cite{Szasz2020}. It is also possible to construct parent hamiltonians having an exact chiral QSL as groundstate, but they are generally long-ranged~\cite{Schroeter2007,Nielsen2013}.

Quite interestingly, it is also possible to realize more exotic FQHE phases with higher spin models, e.g. Moore-Read phase with nonabelian SU(2)$_2$ anyons using spin-1 models~\cite{Chen2018,Liu2018} or Read-Rezayi state with nonabelian SU(2)$_3$ using $S=3/2$ model~\cite{Luo2023}, or even with higher SU($N$) symmetry~\cite{Chen2021}.  Let us also mention that gapped chiral QSL with nonabelian anyons are also found for the Kitaev model (\ref{eq:kitaev2}) in magnetic field~\cite{Zhu2018} or as exact groundstate of a Kitaev model on a decorated triangle-honeycomb lattice~\cite{YaoKivelson2007}.

\subsection{Gapless spectrum}
\subsubsection{Magnetic long-range order}
Although we are interested in disordered phase, for completeness we will review basic properties of magnetically ordered phases. 

When there is a long-range magnetic order due to spontaneous symmetry breaking of the \emph{continuous} spin SU(2) symmetry, there are necesseraly gapless spinwave excitations, corresponding to the Nambu-Goldstone low-energy modes. 

It is also possible to get more involved symmetry breaking, e.g. nematic (or quadrupolar) order. A simple example can be provided in a spin-1 system where the groundstate prefers $S^z=\pm 1$ and not $0$ state locally, i.e. no net magnetization but a preferred direction, hence the name nematic as in liquid crystals, see e.g. a review in~\cite{LacroixMendelsMila_book}.

\subsubsection{Dirac spin liquid}
Using the parton construction described above, it is possible in some cases to assume a static gauge pattern such that the fermionic tight-binding model spectrum exhibits a Dirac spectrum. Then, by enforcing the single-particle occupation constraint, one can construct a spin wavefunction with exotic properties since several physical correlations are algebraic at $T=0$: this state is known as the algebraic Dirac spin liquid (DSL). Its stability in $(2+1)$d is an active field of research since DSL is unstable towards many competing phases~\cite{Hermele2004a} such as magnetic order, chiral QSL or VBC.

Recent numerical studies have argued that DSL could be realized in several microscopic models such as an extended $S=1/2$ triangular Heisenberg model~\cite{Iqbal2016,Hu2019}. Notably, there are specific signatures of low-energy excitations~\cite{Song2020} that could be probed numerically~\cite{Wietek2024} or experimentally. 

In $(3+1)$d, DSL is a stable phase of matter that could describe quantum spin ice~\cite{Gingras2014,review_QSL_2017} with sharp spectroscopic signatures.

\subsubsection{Gapless Z$_2$ QSL}
Going back to Kitaev's model on the honeycomb lattice described in Sec.~\ref{sec:kitaev2}, it also contains a gapless Z$_2$ phase when all couplings are similar in magnitudes. 

In the field theory language, it can be seen as an instability of the DSL by adding a pairing field in the spinon quadratic hamiltonian so that the gauge field symmetry reduces from U(1) to Z$_2$. This could be stabilized on the frustrated $S=1/2$ square lattice~\cite{Hu2013} or on the Shastry-Sutherland lattice~\cite{Yang2022,Viteritti2024}.

Of course, the parton construction is very versatile and many other QSL are possible. For instance, if there is no spinon Fermi surface but only a quadratic band touching, a quadratic gapless Z$_2$ QSL could be stabilized~\cite{Mishmash2013}.

\subsubsection{Spinon Fermi sea}
A natural route to engineer a critical spin wavefunction is to perform a Gutzwiller projection on a filled Fermi sea, using the parton construction.

In 1d, this critical wavefunction is the exact groundstate of the famous Haldane-Shastry spin-1/2 chain with $1/r^2$ interaction~\cite{Haldane1988,Shastry1988} and has a very good variational energy for the usual Heisenberg chain.

In 2d, this wavefunction was proposed as a \emph{spin Bose metal}~\cite{Motrunich2005}. Such a spinon Fermi surface would contain gapless excitations at all momenta in $S({\bf q},\omega)$ and violate area law in the entanglement entropy scaling (as in a Fermi liquid).

\subsubsection{RVB}
Even when considering only nearest-neighbor valence bonds in Fig.~\ref{fig:rvb}, such an RVB wavefunction is nontrivial: clearly, spin correlations are short-range but dimer-dimer correlations decay algebraically~\cite{Albuquerque2010,Tang2011} on a 2d bipartite square lattice, qualitatively similar to the QDM case at its RK point. We have already discussed its parent hamiltonian~\cite{Cano2010} and the fact that in 2d, this can only exist as a critical point. 

The situation is more interesting in 3d where a U(1) QSL phase can be stable on bipartite lattices. For instance, starting from the RK point of a QDM on a bipartite diamond lattice, it is possible to engineer an XXZ model on the dual pyrochlore lattice which could stabilize a U(1) QSL phase~\cite{Hermele2004b} or SU(2) models on decorated lattices with similar properties~\cite{Raman2005}.

\section{Conclusion}
We have described several classical and quantum spin liquids, which by definition do not break any symmetry (neither in spin space or real space). 
Despite this negative definition, we have seen that they can be classified in various categories. 

In the classical case, this classification is rather recent and in progress~\cite{Davier2023,Yan2023}, including the well-known U(1) Coulomb phase (e.g. in classical spin ice materials) but also higher-rank Coulomb phases with fracton-like excitations~\cite{Seiberg2021}. At the moment, the classification mostly relies on the specific form of some models as in Eq.~\ref{eq:H_classical2} and it would be valuable in the future to go beyond and see if a full description of all possible CSL can be obtained. For instance, there are weird CSL analogous to Z$_2$ QSL~\cite{Rehn2017} that deserve further studies.

In the quantum case, there is already a simple distinction depending on the many-body spectrum, see Fig.~\ref{fig:spectrum}, which leads to various qualitative behaviors in some observables, that can be probed experimentally. Moreover, since the seminal work by Wen~\cite{Wen2002}, much progress has been made to classify the different QSL according to the nature of the gauge group and the spinon spectrum. In order to go beyond the PSG classification, braided tensor categories in $(2+1)$d have allowed further progress~\cite{Barkeshli2019} as well as group cohomology~\cite{Liu2024}. As a result, there is a really huge number of distinct QSL and the focus should be now to understand on which microscopic models they can be realized. It would be crucial to understand if this classification is valid beyond mean-field, and in particular connect the solutions that can be found using a bosonic vs fermionic slave-particle description.

Thanks to some famous exact solutions that we have partly reviewed, various gapped or gapless QSL are known to be stable in 2d or 3d, which pave the way for their realization using realistic microscopic models or existing quantum materials. 
We have discussed in details some famous examples of QSL such as gapped Z$_2$, gapless U(1), chiral spin liquid, Dirac spin liquid, spin Bose metal\ldots 
Among these QSL, a lot has still to be understood particularly on the Dirac spin liquid in $(2+1)$d which may or may not be stable as a phase.

Besides the achievements of analytical tools, there is a large activity in numerical simulations with the developments of novel algorithms (e.g. tensor networks, neural networks etc.). While we focused on zero-temperature, it is crucial to further improve these methods to tackle finite-temperature (for thermodynamics) as well as finite-time evolution (for dynamics). These numerical tools are called for since there are very few exact results and one would like to investigate realistic models for classical or quantum spin liquids.

While we have focused on describing some exotic phases of matter, it is also very important to understand whether there can exist unconventional phase transitions between them, which description is beyond the standard Landau-Ginzburg paradigm. This can occur for instance between a magnetically ordered phase and a valence-bond crystal one, so-called deconfined quantum criticality~\cite{Senthil2023} or weakly first-order behavior~\cite{Takahashi2024}, or even more exotic Stiefel liquids which have no Lagrangian descriptions~\cite{Zou2021} that could describe quantum critical spin liquids.

\smallskip
{\bf Acknowledgment: }
 I would like to thank Yvan Castin and Carlos S\'a de Melo for the organisation of the symposium "Open questions in the quantum many-body problem" at Institut Henri Poincar\'e (IHP, Paris)  where this lecture was presented.

% The next command determines the bibliography style. Please do not
% change this.

\bibliographystyle{crunsrt}
\bibliography{ref}
\end{document}